\documentclass[aps,prd,amsmath,amssymb,showpacs,10pt,superscriptaddress]{revtex4-2}
\usepackage{graphicx}
\usepackage{color}
\usepackage{slashed}
\usepackage{subfigure}
\usepackage{enumerate}
\usepackage{physics}
\usepackage[font=footnotesize]{caption}
\usepackage[breaklinks,colorlinks,urlcolor=blue,linkcolor=red,anchorcolor=magenta,citecolor=blue]{hyperref}
\usepackage{cleveref} 
\crefname{table}{Table}{Tables}
\crefname{equation}{Eq.}{Eqs.}
\crefname{figure}{Fig.}{Figs.}
\crefname{section}{Sec.}{Secs.}
\allowdisplaybreaks
\begin{document}
\title{Combined analysis of the singly-Cabbibo-suppressed decays of $D^{0} \to VP$}
\author{Jun Wang
}\email{junwang@ihep.ac.cn}
\affiliation{Institute of High Energy Physics, Chinese Academy of Sciences, Beijing 100049, People's Republic of China}
\affiliation{University of Chinese Academy of Sciences, Beijing 100049, People's Republic of China}
\author{Qiang Zhao
}\email{ zhaoq@ihep.ac.cn} 
\affiliation{Institute of High Energy Physics, Chinese Academy of Sciences, Beijing 100049, People's Republic of China}
\affiliation{University of Chinese Academy of Sciences, Beijing 100049, People's Republic of China}
\affiliation{Center for High Energy Physics, Henan Academy of Sciences, Zhengzhou 450046, People's Republic of China}

\begin{abstract}
We investigate six singly Cabibbo-suppressed decay channels in $D^0\to VP$ ( $V$ and $P$ stand for the ground state vector and pseudoscalar mesons, respectively), i.e.  $D^{0}\to \rho^{+}\pi^{-}$, $\rho^{-}\pi^{+}$, $K^{*+}K^{-}$, $K^{*-}K^{+}$, $K^{*0}\bar{K}^{0}$, and $\bar{K}^{*0}K^{0}$. These decay channels share the similar transition mechanisms involving only the direct emission (DE) and internal conversion (IC) processes. We show that a combined analysis of these channels can explicitly highlight the role played by the IC processes which contribute to the amplitudes at the same order of magnitude as the DE processes.
\end{abstract}
\maketitle

\section{Introduction}
For $B$ mesons, due to the sufficiently heavy mass of the $b$ quark, QCD-inspired approaches can well describe the two-body non-leptonic decays of $B$ mesons~\cite{Chernyak:1983ej,Chernyak:1981zz,
Brodsky:1981kj,Nayak:2024esq,Egner:2024azu,Mannel:2024uar,Sheng:2025qun,Mannel:2025fvj,Wang:2025rtg}. However, for $D$ mesons, since the charm quark mass is $m_{c}\sim 1.5~\mathrm{GeV}$, which lies at the boundary between the perturbative and non-perturbative regimes, non-perturbative effects are expected to play a significant role~\cite{Cheng:2021yrn,Cheng:2010ry,
Cheng:2019ggx,Cao:2023csx,Cao:2023gfv}. In recent years, with the improvement of experimental precision, abundant experimental data on the two-body non-leptonic decays of $D^0$ mesons have been accumulated~\cite{ParticleDataGroup:2024cfk,BESIII:2023exz,dArgent:2017gzv}. Based on these data, a systematic study of $D^0$ weak decays can be performed, which facilitates a deeper understanding of non-perturbative interactions in the charm energy region. Furthermore, $D^0$ weak decays are often accompanied by intermediate strong interaction contributions. Therefore, investigating these decays can also help clarify strong interaction mechanisms, such as those involving exotic hadrons~\cite{Rahmani:2025uut,Chen:2022asf,Liu:2019zoy,Zhu:2004xa,Chen:2016qju,Liu:2024uxn}.

The study of $D^{0}$ weak decays is primarily conducted using the topological diagram approach (TDA)~\cite{Li:2012cfa,Qin:2013tje,Cheng:2012xb,Cheng:2012wr,Cheng:2016ejf, Cheng:2019ggx,Cheng:2010ry,Cheng:2021yrn,Chau:1986jb,Cheng:2010rv, Cheng:2024hdo}, where the contributions from different diagrams are extracted through global fits to experimental data. During the past few years a quark model approach has been developed and applied to charmed hadron weak decays~\cite{Cao:2023csx,Cao:2023gfv,Niu:2020aoz,Niu:2020gjw,Niu:2021qcc,Niu:2025lgt}. The advantage is that the wave function convolution can provide not only a natural form factor based on the quark model phenomenology, but also a unified way for taking into account the SU(3) flavor symmetry breaking. The transition amplitudes calculated in the quark model can also provide an explanation for the dynamic origin of the topological diagram approach. It is also useful that by examining the quark model relations among these channels which can be connected with each other by SU(3) flavor symmetry, apparent deviations will indicate non-trivial mechanisms for the underlying dynamics. For instance, in Ref.~\cite{Cao:2023csx} it was shown that the significant difference between $D^0\to \phi\omega$ and $\phi\rho^0$ has provided a clear evidence for other non-perturbative mechanisms beyond the leading-order quark model expectation.

In this work, we focus on six singly Cabibbo-suppressed channels in $D^{0}\to VP$, i.e., $D^{0}\to \rho^{+}\pi^{-}$, $\rho^{-}\pi^{+}$, $K^{*+}K^{-}$, $K^{*-}K^{+}$, $K^{*0}\bar{K}^{0}$, and $\bar{K}^{*0}K^{0}$. We consider the leading-order short-distance contributions, specifically the color-allowed direct emission (DE) and the color-suppressed internal conversion (IC) induced by the $W$-exchange between the quark and anti-quark within the $D^{0}$ meson. While the DE process can be reliably calculated in the quark model framework, we will discuss the consequence of the DE dominance, which will call for the contributions from the IC process. Note that the decay channels of $D^0\to K^{*0}\bar{K}^{0}$, and $\bar{K}^{*0}K^{0}$ only involve the IC transitions as the leading-order short-distance contributions. Thus, the experimental data will provide constraints on the IC transition. We will show that the IC plays an indispensable role in $D^0\to VP$.

The remainder of this paper is organized as follows. We first introduce our formalism in \cref{sec:formalism}. The numerical results and discussions are presented in \cref{sec:results}. A brief summary and conclusion are given in \cref{sec:summary}.

\section{Formalism}\label{sec:formalism}
\begin{figure}[!h]
    \centering
     \subfigure[$D^{0}\to \rho^{\pm}\pi^{\mp}$ DE process]{
        \includegraphics[width=0.4\textwidth]{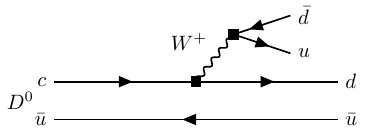}\label{fig:rhopiDE}
    }
      \subfigure[$D^{0}\to \rho^{\pm}\pi^{\mp}$ IC process]{
        \includegraphics[width=0.4\textwidth]{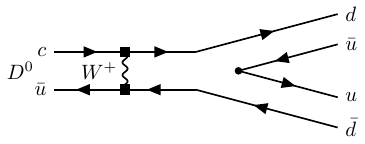}\label{fig:rhopiIC}
    }
     \subfigure[$D^{0}\to K^{*\pm}K^{\mp}$ DE process]{
        \includegraphics[width=0.4\textwidth]{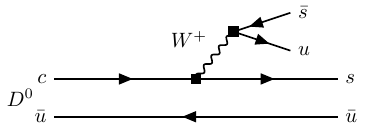}\label{fig:KstarKDE}
    }
      \subfigure[$D^{0}\to K^{*\pm}K^{\mp}$ IC process]{
        \includegraphics[width=0.4\textwidth]{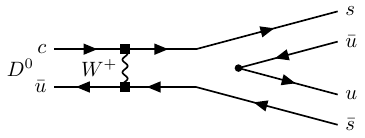}\label{fig:KstarKIC}
    }
     \subfigure[$D^{0}\to K^{(*)0}\bar{K}^{(*)0}$ IC process ($s \bar{s}$)]{
        \includegraphics[width=0.4\textwidth]{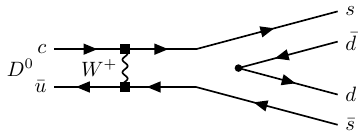}\label{fig:KstarKIC1}
    }
      \subfigure[$D^{0}\to K^{(*)0}\bar{K}^{(*)0}$ IC process ($d \bar{d}$)]{
        \includegraphics[width=0.4\textwidth]{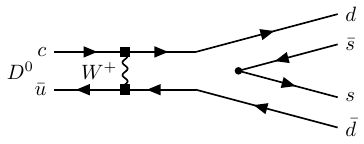}\label{fig:KstarKIC2}
    }
    \caption{Schematic diagrams for $D^{0}\to VP$ processes considered in this work.}   \label{fig:DEIC}
  \end{figure}

We consider six singly Cabibbo-suppressed channels in $D^{0}\to VP$: $D^{0}\to \rho^{+}\pi^{-}$, $\rho^{-}\pi^{+}$, $K^{*+}K^{-}$, $K^{*-}K^{+}$, $K^{*0}\bar{K}^{0}$, and $\bar{K}^{*0}K^{0}$. For these six channels, the amplitudes are decomposed into direct emission (DE) and internal conversion (IC) contributions, as illustrated in \cref{fig:DEIC}. In particular, both DE and IC contribute to the $\rho^{\pm}\pi^{\mp}$ and $K^{*\pm}K^{\mp}$ channels, whereas the $K^{(*)0}\bar{K}^{(*)0}$ channels receive contributions only from the IC. This feature not only provides evidences for the role played by the IC mechanism, but also sets up constraints on the magnitude of the IC. 

\subsection{Direct emission (DE) transition}
   At the quark level, the DE process corresponds to a $1\to 3$ decay of the initial charm quark. The relevant operator is $\hat{H}^{(P)}_{W,1\to 3}$, and for the $D^{0} \to VP$ process, only the parity-conserving part contributes. The explicit form of the operator can be found in Refs.~\cite{LeYaouanc:1988fx,Niu:2020gjw}. The corresponding amplitude is given by
  \begin{equation}
     \mathcal{M}^{q}_{\text{DE}}(D^{0}\to V P)=\mel*{P(\mathbf{P}_{3})V(\mathbf{P}_{2})}{\hat{H}_{1\to 3}}{D^{0}(\mathbf{P}_{1})}
     \end{equation}
The above amplitudes are calculated using the non-relativistic constituent quark model (NRCQM) wave functions~\cite{Kokoski:1985is,Godfrey:1985xj,Godfrey:1986wj,Niu:2025lgt,Niu:2020gjw}.
Due to the normalization conventions, the amplitude at the quark level differs from that at the hadronic level by a factor. The corresponding hadronic-level amplitude is given by
 \begin{equation}
   \mathcal{M}^{h}_{\text{DE}}=-8\pi^{\frac{3}{2}}\sqrt{m_{D^{0}}E_{V}E_{P}}\mathcal{M}^{q}_{\text{DE}}\,.
  \end{equation} 
Since $D^{0}(p_1)\to V(\varepsilon_{2},p_{2})P(p_3)$ is a $P$-wave decay, we   define the relevant effective coupling constants as
\begin{equation}
 \mathcal{M}^{h}_{\text{DE}}(D^{0}\to VP)\equiv V_{cq}V_{uq}\mathcal{G}^{\text{DE}}\varepsilon_{2}\cdot p_3,\,
 \end{equation} 
where $V_{cq}$ and $V_{uq}$ are the relevant Cabibbo-Kobayashi-Maskawa (CKM) matrix elements $q=d,s$.

 Substituting the corresponding wave functions and operators into the above expressions, we obtain the corresponding effective coupling constants for the DE processes as
         \begin{eqnarray}
        \mathcal{G}^{\text{DE}}_{\rho^{+}\pi^{-}}&=& \frac{2 G_F m_{\rho }  (R_{\pi } R_D R_{\rho })^{3/2} e^{-\frac{\abs{\mathbf{p}}^2}{8 (R_D^2+R_{\pi }^2)}}\sqrt{E_\pi E_\rho m_{D^0}} }{\pi ^{3/4} m_c m_{D^0} (R_D^2+R_{\pi }^2)^{5/2} m_q^3}\nonumber\\
       &&\times (2 m_c (R_D^2+R_{\pi }^2) m_q^2+m_c (R_D^2+2 R_{\pi }^2) m_q^2-R_D^2 m_q^3) ,\label{rho+}\\
       \mathcal{G}^{\text{DE}}_{K^{*+}K^{-}}&=& \frac{2 G_F m_{K^*} (R_D R_K R_{K^*})^{3/2} e^{-\frac{\abs{\mathbf{p}}^2}{8 (R_D^2+R_K^2)}} \sqrt{E_K E_{K^*} m_{D^0}} }{\pi ^{3/4} m_c m_{D^0} m_q m_s^2 (R_D^2+R_K^2)^{5/2}}\nonumber\\
       &&\times (m_c m_s (R_D^2+R_K^2) (m_q+m_s)+m_c m_q m_s (R_D^2+2 R_K^2)-R_D^2 m_q m_s^2),\label{K+} \\ 
  \mathcal{G}^{\text{DE}}_{\rho^{-}\pi^{+}}&=&\mathcal{G}^{\text{DE}}_{\rho^{+}\pi^{-}}(R_{\rho}\leftrightarrow R_{\pi}),\label{rho-}\\ \mathcal{G}^{\text{DE}}_{K^{*-}K^{+}}&=&\mathcal{G}^{\text{DE}}_{K^{*+}K^{-}}(R_{K^*}\leftrightarrow R_{K})\,,\label{K-}
 \end{eqnarray}
where $m_q$ and $m_s$ are the constituent quark masses for the $u/d$ and $s$ quarks, respectively. $R_{D}$, $R_{\pi}$, $R_{\rho}$, $R_{K}$, and $R_{K^{*}}$ are the harmonic-oscillator strengths of the corresponding mesons. $\abs{\mathbf{p}}$ is the magnitude of the three-momentum of the final-state mesons in the $D^{0}$ rest frame, and $E_{\pi}$, $E_{\rho}$, $E_{K}$, and $E_{K^{*}}$ are their energies. $G_{F}$ is the Fermi coupling constant.

\subsection{Internal conversion (IC) transition}
The IC process corresponds to a $2\to 2$ scattering of the initial $c$ quark and $\bar{u}$ quark, producing intermediate states, such as $\pi^0$ and $\pi^{0*}$ in $D^0\to \rho^{\pm}\pi^\mp$ via the $d\bar{d}$ component, and both $\pi^{0(*)}$ via the $d\bar{d}$ component and  $\eta^{(*)}/\eta^{\prime(*)}$ via the $s\bar{s}$ component in $D^0\to K^{(*)0}\bar{K}^{(*)0}$, which then couple to $VP$ through strong interactions. In principle, all the intermediate states with $J^{PC}=0^{-+}$ and correct $G$-parity should contribute to the IC amplitudes. 

Taking into account that many states with $J^{PC}=0^{-+}$ have not yet been well established in experiment, explicit inclusion of the intermediate states here will be strongly model-dependent. Hence, we adopt a slightly different strategy to leave the IC to be constrained by the experimental data. We will then discuss the behavior of the intermediate pole terms.  Due to the width effects from the intermediate states,  the IC contributions can be defined by a complex parameter $\mathcal{G}^{\text{IC}}$. The total amplitude can then be written as
\begin{equation}
  \mathcal{M}(D^{0}\to V(\varepsilon_{2},p_{2})P(p_{3}))= V_{cq}V_{uq} (\mathcal{G}^{\text{DE}}+\mathcal{G}^{\text{IC}})\varepsilon_{2}\cdot p_{3}\equiv \mathcal{G}^{\text{total}}\varepsilon_{2}\cdot p_{3} \,.\end{equation}
As mentioned earlier in the IC process, $D^{0}\to \rho^{\pm}\pi^{\mp}$ proceeds via coupling to $d\bar{d}$, $D^{0}\to K^{*\pm}K^{\mp}$ via coupling to $s\bar{s}$, and $D^{0}\to K^{(*)0}\bar{K}^{(*)0}$ can couple to both $s\bar{s}$ and $d\bar{d}$. Therefore we parameterize the coupling  based on $SU(3)$ flavor symmetry. We assume that the IC contributions from $d\bar{d}$ in the $\rho^{\pm}\pi^{\mp}$ channel and from $s\bar{s}$ in the $K^{*\pm}K^{\mp}$ channel are identical to those in the $K^{(*)0}\bar{K}^{(*)0}$ channel. Therefore, the effective coupling constants for these channels can be written as
\begin{equation}
\begin{aligned}\label{eq:amp_1}
\mathcal{G}^{\text{total}}(D^{0}\to \rho^{\pm}\pi^{\mp})=&V_{cd}V_{ud}(\mathcal{G}^{\text{DE}}_{\rho^{\pm}\pi^{\mp}}+\mathcal{G}^{\text{IC}}_{d\bar{d}}) \,, \quad \mathcal{G}^{\text{IC}}_{d \bar{d}}=\abs{\mathcal{G}^{\text{IC}}_{d \bar{d}}}e^{i \theta_{d \bar{d}}}\\\mathcal{G}^{\text{total}}(D^{0}\to K^{*\pm}K^{\mp})=&V_{cs}V_{us}(\mathcal{G}^{\text{DE}}_{K^{*\pm}K^{\mp}}+\mathcal{G}^{\text{IC}}_{s \bar{s}})\,, \quad \mathcal{G}^{\text{IC}}_{s \bar{s}}=\abs{\mathcal{G}^{\text{IC}}_{s \bar{s}}}e^{i \theta_{s \bar{s}}}\\
  \mathcal{G}^{\text{total}}(D^{0}\to K^{(*)0}\bar{K}^{(*)0})=&V_{cs}V_{us}\mathcal{G}^{\text{IC}}_{s \bar{s}}+e^{i \theta_{d s}}V_{cd}V_{ud}\mathcal{G}^{\text{IC}}_{d \bar{d}}\,,
\end{aligned}
 \end{equation} 
 where $\theta_{ds}$ is the relative phase between the $s\bar{s}$ and $d\bar{d}$ contributions in the $D^{0}\to K^{(*)0}\bar{K}^{(*)0}$ process.

With the above effective couplings the total decay width for these considered channels can be obtained by
\begin{equation}
  \Gamma(D^{0}\to VP)=\frac{\abs{\mathbf{p}}^{3}}{8\pi m_{V}^{2}}\abs{\mathcal{G}^{\text{total}}}^{2} \,.
 \end{equation}

\section{Numerical Results and Discussions}\label{sec:results}
In \cref{tab:hos}, we list the values of the relevant coupling constants and parameters used in the NRCQM, which are commonly adopted in the study of hadron spectra~\cite{Godfrey:1985xj,Kokoski:1985is,Godfrey:2016nwn,Godfrey:2004ya,Godfrey:1986wj,Godfrey:2015dva}.

  \begin{table}[!h]
      \centering\caption{Harmonic oscillator (HO) strengths for meson wave functions and the constituent quark masses \cite{Godfrey:1985xj,Kokoski:1985is,Godfrey:2016nwn,Godfrey:2004ya,Godfrey:1986wj,Godfrey:2015dva}.}
      \begin{ruledtabular}
        \begin{tabular}{cccc}
          HO Strength&Values ($\mathrm{MeV}$)& Quark mass & Values ($\mathrm{MeV}$)\\\hline
          $R_{D}$&660&$m_{c}$&$1628$\\
         $R_{K^{*}}$&480 &$m_{s}$&$419$\\
        $R_{K}$ &710 &$m_{u}$&$220$\\
        $R_{\rho}$&450&$m_{d}$&$220$\\
        $R_{\pi}$&750&&\\
             \end{tabular}\label{tab:hos}
      \end{ruledtabular}
       \end{table}

In \cref{tab:GDE}, we present the numerical results for the effective coupling constants $\mathcal{G}^{\text{DE}}$. These are the theoretical predictions for the DE transition based on the NRCQM. With $\mathcal{G}^{\text{DE}}$ as input the experimental branching ratios are fitted, and the fitted parameters are listed in \cref{tab:fit_para}. The fitting results for the branching ratios and the corresponding experimental data are listed in \cref{tab:fit_BR}.

  \begin{table}[!h]
      \centering\caption{Numerical results of the effective coupling constants $\mathcal{G}^{\text{DE}}$ for the direct emission processes.}
      \begin{ruledtabular}
        \begin{tabular}{ccccc}
         Decay channels &$D^{0}\to\rho^{+}\pi^{-}$&$D^{0}\to\rho^{-}\pi^{+}$&$D^{0}\to K^{*+}K^{-}$&$D^{0}\to K^{*-}K^{+}$\\\hline
         $\mathcal{G}^{\text{DE}}(\times 10^{-5})$&0.805&1.391&0.683&1.059
             \end{tabular}\label{tab:GDE}
      \end{ruledtabular}
       \end{table}

\begin{table}[!ht]
    \centering\caption{Fitted values of the parameters.}\label{tab:fit_para}
      \begin{ruledtabular}
      \begin{tabular}{ccccccc}
       Parameters&$\abs{\mathcal{G}^{\text{IC}}_{d \bar{d}}}(\times 10^{-5})$&$\theta_{d \bar{d}}(^\circ)$&$\abs{\mathcal{G}^{\text{IC}}_{s \bar{s}}}(\times 10^{-5})$&$\theta_{s \bar{s}}(^\circ)$&$\theta_{ds}(^\circ)$&$\chi^{2}/\text{d.o.f}$\\\hline
Fit$-1$&$1.146\pm 0.004$&$180.0\pm 2.3$&$1.053\pm 0.007$&$167.5\pm 0.7$&$192.5\pm 3.7$&2.3\\
Fit$-2$&$1.146\pm 0.004$&$\setminus$&$1.053\pm 0.007$&$167.5\pm 0.7$&$\setminus$&0.7
\end{tabular}
  \end{ruledtabular}
\end{table}

\begin{table}[!h]
      \centering\caption{The fitted branching ratios for the relevant decay channels in two different fitting schemes. The experimental data from PDG~\cite{ParticleDataGroup:2024cfk} are also included for a comparison.}
      \begin{ruledtabular}
        \begin{tabular}{ccccccc}
         Decay channels &$D^{0}\to\rho^{+}\pi^{-}$&$D^{0}\to\rho^{-}\pi^{+}$&$D^{0}\to K^{*+}K^{-}$&$D^{0}\to K^{*-}K^{+}$&$D^{0}\to K^{*0}\bar{K}^{0}$&$D^{0}\to \bar{K}^{*0}K^{0}$\\\hline
         $\mathcal{B}_{\text{exp}}(\times 10^{-3})$&$10.1\pm 0.5 $&$5.15\pm 0.26$&$5.36\pm 0.61$&$1.82\pm 0.20$&$0.342\pm 0.063$&$0.249\pm 0.048$\\
         $\mathcal{B}_{\text{fit$-1$}}(\times 10^{-3})$&$10.3\pm 0.3$&$5.27\pm 0.21$&$5.80\pm 0.27$&$1.79\pm 0.19$&$0.290\pm 0.139 $&$0.290\pm 0.139 $\\
           $\mathcal{B}_{\text{fit$-2$}}(\times 10^{-3})$&$10.3\pm 0.3$&$5.27\pm 0.21$&$5.80\pm 0.27$&$1.79\pm 0.19$&$0.290\pm 0.055 $&$0.290\pm 0.055 $
             \end{tabular}\label{tab:fit_BR}
      \end{ruledtabular}
       \end{table}

In Fit$-1$, we use five parameters to fit six decay channels. For $D^{0}\to \rho^{\pm}\pi^{\mp}$, the phase of $\mathcal{G}^{\text{IC}}_{d \bar{d}}$, $\theta_{d \bar{d}}$ is found to favor a destructive phase with $\theta_{d \bar{d}}=(180.0\pm 2.3)^{\circ}$,  indicating that the IC transition via the intermediate states containing a $d\bar{d}$ component will interfere destructively with the DE amplitude. In $D^{0}\to K^{(*)0}\bar{K}^{(*)0}$, the fitted relative phase between $\mathcal{G}^{\text{IC}}_{d \bar{d}}$ and $\mathcal{G}^{\text{IC}}_{s \bar{s}}$ is $192.5^{\circ}$, which is also a destructive phase between these two transitions. Note that here the intermediate $q\bar{q}$ can have quantum numbers of both $I^G J^{PC}=0^+0^{-+}$ and $I^G J^{PC}=1^-0^{-+}$. The $I^G=0^+$ contribution seems to be consistent with the expectation from the intermediate $\eta$ state based on SU(3) flavor symmetry \cite{Cao:2023csx,Cao:2023gfv}, since the intermediate $\eta$ state coupling to both $d\bar{d}$ and $s\bar{s}$ naturally leads to a destructive phase. The $I^G=1^-$ component from the intermediate $d\bar{d}$ has a similar behavior as that in $D^0\to \rho^\pm\pi^\mp$.

For the decay channels considered in this work, we find that the contributions from the DE and IC transitions are comparable. This is consistent with the conclusion obtained from global fits using the TDA~\cite{Cheng:2024hdo,Zheng:2025ryf,Cheng:2024hdo,Cheng:2010ry,Cheng:2019ggx,Cheng:2016ejf,Zhong:2004ck,Cheng:2012wr,Qin:2013tje,Niu:2025lgt,Jiang:2017zwr} (see~\cref{tab:compare}). However, our results indicate that the DE and IC amplitudes are nearly out of phase, leading to destructive interferences, whereas the TDA fit yields a relative phase close to $90^\circ$. Another different feature is that the DE amplitude of $D^0\to \rho^+\pi^-$  in the NRCQM is smaller than that of $D^0\to \rho^-\pi^+$. In the TDA the relative magnitude of these two amplitudes is opposite as shown in ~\cref{tab:compare}. 

Actually, the relative magnitude of the DE amplitudes between these two charged channels, i.e. $D^0\to V^+P^-$ and $V^-P^+$,  is fixed in the NRCQM. 
Analyzing the analytical expressions for the DE process, we find that $\mathcal{G}_{V^{-}P^{+}}^{\text{DE}} > \mathcal{G}_{V^{+}P^{-}}^{\text{DE}}$, primarily due to the relatively large harmonic oscillator parameters chosen for the pseudoscalar mesons. As shown by Eqs.~(\ref{rho+})-(\ref{K-}), the relative coupling strength between the $V^+P^-$ and $V^-P^+$ channel is given by the following relation:
 \begin{equation}
  \quad \mathcal{G}^{\text{DE}}_{V^{+}P^{-}}\propto \frac{3 R_{D}^{2}+4R_{P}^{2}}{(R_{D}^{2}+R_{P}^{2})^{\frac{5}{2}}},\quad \mathcal{G}^{\text{DE}}_{V^{-}P^{+}}\propto \frac{3 R_{D}^{2}+4R_{V}^{2}}{(R_{D}^{2}+R_{V}^{2})^{\frac{5}{2}}},\quad \frac{\mathcal{G}^{\text{DE}}_{V^{-}P^{+}}}{\mathcal{G}^{\text{DE}}_{V^{+}P^{-}}}\simeq 1.5 \label{eq:DEcom}
   \end{equation} 
The relatively large harmonic oscillator parameters chosen for the pseudoscalar mesons
agree with the LQCD results for the light-cone distribution amplitudes of pseudoscalar mesons, which exhibit a broader distribution in momentum space~\cite{LatticeParton:2022zqc}. They are also the values favored by the NRCQM calculations of the meson spectra~\cite{Godfrey:1985xj,Kokoski:1985is,Godfrey:2016nwn,Godfrey:2004ya,Godfrey:1986wj,Godfrey:2015dva}. 
As a consequence of $\mathcal{G}_{V^{-}P^{+}}^{\text{DE}} > \mathcal{G}_{V^{+}P^{-}}^{\text{DE}}$ in the NRCQM, supposing that only the DE transition contributes in $D^0\to \rho^\pm\pi^\mp$ (or $K^{*\pm}K^{\mp}$), the NRCQM predictions for $\rho^\pm\pi^\mp$ ($K^{*\pm}K^{\mp}$) will have reversed relative magnitude in comparison with the experimental data~\cite{ParticleDataGroup:2024cfk}. This is an indication that additional contributions are needed. In this case, it naturally refers to the IC transition in these decay channels. Comparing the fitted values for couplings $\abs{\mathcal{G}^{\text{IC}}_{d \bar{d}}}$ and $\abs{\mathcal{G}^{\text{IC}}_{s \bar{s}}}$ in ~\cref{tab:fit_para} with the calculated coupling values for the DE transitions, we find that they are comparable with each other.

To better understand the role played by the IC transition, one may first understand whether the NRCQM provides a reasonable description of the DE process. Namely, whether $\mathcal{G}_{V^{-}P^{+}}^{\text{DE}} > \mathcal{G}_{V^{+}P^{-}}^{\text{DE}}$ in the NRCQM is acceptable or not. 
The behavior of the DE transition can be tested through the semileptonic decays of $D^{0}$. We consider the processes $D^{0}\to \rho^{-}e^{+}\nu_{e}$ and $D^{0}\to \pi^{-}e^{+}\nu_{e}$, which correspond to the weak transition form factors of $D^{0}\to \rho^-$ and $D^{0}\to \pi^-$, respectively. They can be compared with the relative magnitudes of $\mathcal{G}_{\rho^{-}\pi^{+}}^{\text{DE}}$ and $\mathcal{G}_{\rho^{+}\pi^{-}}^{\text{DE}}$. The measured branching ratios~\cite{ParticleDataGroup:2024cfk} are
\begin{equation}
  \mathcal{B}(D^{0}\to \rho^{-}e^{+}\nu_{e})=(1.46\pm 0.08)\times 10^{-3},\quad \mathcal{B}(D^{0}\to \pi^{-}e^{+}\nu_{e})=(2.91\pm 0.04)\times 10^{-3}\,.
 \end{equation} 
After taking off the phase space effects, we obtain
\begin{equation}
  \abs{\frac{\mathcal{M}(D^{0}\to \rho^{-}e^{+}\nu_{e})}{\mathcal{M}(D^{0}\to \pi^{-}e^{+}\nu_{e})}}\simeq 1.2,
 \end{equation} 
which agrees with the relation $\mathcal{G}_{\rho^{-}\pi^{+}}^{\text{DE}} > \mathcal{G}_{\rho^{+}\pi^{-}}^{\text{DE}}$. It means that the relative strengths of the DE amplitudes between the $V^+P^-$ and $V^-P^+$ are correctly dealt with in the NRCQM.

To test the sensitivity of the DE contribution to the wave function parameters, we modify the harmonic oscillator parameters for $K$ and $\pi$ mesons to $R_{\pi}=700~\mathrm{MeV}$ and $R_{K}=670~\mathrm{MeV}$, respectively. These values are within the ranges for the parameters which can be found in the literature.  The amplitudes are then fitted in the same manner as in Fit$-1$. The fitting results and branching ratios are presented in \cref{tab:fit_para_t,tab:fit_BR_t}. From the results, we observe that the outcomes remain consistent, indicating that the DE contribution is insensitive to the harmonic oscillator parameters.

 \begin{table}[!ht]
    \centering
    \caption{Couplings and fitted phase angles compared with the those extracted by the TDA.
     The magnitudes of the coupling constants are quoted in units of $10^{-6}$, and the phase angle are presented in degrees $(^{\circ})$.}
    \label{tab:compare}
     \begin{ruledtabular}
    \begin{tabular}{cccccccc}
      Parameters&$\mathcal{G}^{\text{DE}}_{\rho^{+}\pi^{-}}(\abs{T_{P}}$)&$\setminus(\delta_{T_{P}})$&$\mathcal{G}^{\text{IC}}_{\rho^{+}\pi^{-}}(\abs{E_{V}})$&$\theta_{d \bar{d}}(\delta_{E_{V}})$&$\mathcal{G}^{\text{DE}}_{\rho^{-}\pi^{+}}(\abs{T_{V}}) $&$\mathcal{G}^{\text{IC}}_{\rho^{-}\pi^{+}}(\abs{E_{P}})$&$\theta_{d \bar{d}}(\delta_{E_{P}})$\\\hline
Our works&$8.05$&$\setminus $&$11.46\pm 0.04$&$180.0\pm 2.3$&$13.91$&$11.46$&$180.0\pm 2.3$\\
TDA&$3.58\pm 0.06$&${327^{+5}_{-4}}$&$0.92\pm 0.04$&$92\pm 2$&$2.17\pm 0.03$&$ 1.65 \pm 0.03$&$253\pm 3 $\\

\end{tabular}
 \end{ruledtabular}
\end{table}

Note that in Fit$-1$, our results indicate that $\theta_{d\bar{d}} = 180^\circ$ with a small uncertainty, and the value of $\theta_{ds}$ ensures that $\mathcal{G}^{\text{IC}}_{d \bar{d}}$ and $\mathcal{G}^{\text{IC}}_{s \bar{s}}$ are nearly out of phase, leading to destructive interference between these two complex amplitudes. To further investigate the role played by the IC transition and its interference with the DE amplitudes, we carry out Fit$-2$ (see \cref{tab:fit_para}) by directly setting $\mathcal{G}^{\text{IC}}_{d \bar{d}}$ to be real and out of phase with $\mathcal{G}^{\text{DE}}_{\rho^{\pm}\pi^{\mp}}$. In the $K^{(*)0}\bar{K}^{(*)0}$ channel, we also set $\mathcal{G}^{\text{IC}}_{s \bar{s}}$ to be out of phase with $\mathcal{G}^{\text{IC}}_{d \bar{d}}$. That is, the relevant coupling constants in Eq.~\eqref{eq:amp_1} are replaced by those shown in Eq.~\eqref{eq:amp_2} below, and the number of parameters is reduced to 3: 
\begin{equation}
\label{eq:amp_2}
\mathcal{G}^{\text{total}}(D^{0}\to \rho^{\pm}\pi^{\mp})=V_{cd}V_{ud}(\mathcal{G}^{\text{DE}}_{\rho^{\pm}\pi^{\mp}}-\abs{\mathcal{G}^{\text{IC}}_{d\bar{d}}}) \,, \quad 
  \mathcal{G}^{\text{total}}(D^{0}\to K^{(*)0}\bar{K}^{(*)0})=V_{cs}V_{us}\abs{\mathcal{G}^{\text{IC}}_{s \bar{s}}}-V_{cd}V_{ud}\abs{\mathcal{G}^{\text{IC}}_{d \bar{d}}}\,.
 \end{equation} 
As shown in \cref{tab:fit_para}, the fitted parameters in Fit$-1$ and Fit$-2$ are almost identical. However, due to the reduction in the number of free parameters, the fitted value of $\chi^{2}/\text{d.o.f}$ decreases from $2.3$ in Fit$-1$ to $0.7$ in Fit$-2$.

From the fitted branching ratios (see \cref{tab:fit_BR}), the central values obtained from the two fitting schemes are identical.  In Fit$-1$, the uncertainty of the fitted branching ratio for $D^{0}\to K^{(*)0}\bar{K}^{(*)0}$ is relatively large, since all five free parameters are involved in this channel. In contrast, in Fit$-2$, only two free parameters are relevant for $D^{0}\to K^{(*)0}\bar{K}^{(*)0}$, which significantly reduces the uncertainty of the fitted branching ratio. In particular, one notices that these two channels only involve the IC transition as the leading mechanism, the success of the combined analysis also shows that the IC amplitudes have not been overestimated. In other words, the yields of the DE transition from the NRCQM, despite having a destructive phase to the IC ones, provided a reasonable description of the magnitudes of the DE amplitudes.

\begin{table}[!ht]
    \centering\caption{Fitted values of the parameters with different $R_{\pi}$ and $R_{K}$.}\label{tab:fit_para_t}
      \begin{ruledtabular}
      \begin{tabular}{ccccccc}
       Parameters&$\abs{\mathcal{G}^{\text{IC}}_{d \bar{d}}}(\times 10^{-5})$&$\theta_{d \bar{d}}(^\circ)$&$\abs{\mathcal{G}^{\text{IC}}_{s \bar{s}}}(\times 10^{-5})$&$\theta_{s \bar{s}}(^\circ)$&$\theta_{ds}(^\circ)$&$\chi^{2}/\text{d.o.f}$\\\hline
Fit$-3$&$1.102\pm 0.007$&$170.8\pm 0.4$&$1.028\pm 0.029$&$167.1\pm 1.1$&$180.8\pm 3.1$&$1.4$
\end{tabular}
  \end{ruledtabular}
\end{table}

\begin{table}[!h]
      \centering\caption{Fitted branching ratios for the relevant decay channels with different $R_{\pi}$ and $R_{K}$.}
      \begin{ruledtabular}
        \begin{tabular}{ccccccc}
         Decay channels &$D^{0}\to\rho^{+}\pi^{-}$&$D^{0}\to\rho^{-}\pi^{+}$&$D^{0}\to K^{*+}K^{-}$&$D^{0}\to K^{*-}K^{+}$&$D^{0}\to K^{*0}\bar{K}^{0}$&$D^{0}\to \bar{K}^{*0}K^{0}$\\\hline
         $\mathcal{B}_{\text{fit$-3$}}(\times 10^{-3})$&$10.1\pm 0.5$&$5.15\pm 0.3$&$5.36\pm 0.7$&$1.82 \pm 0.3$&$0.28 \pm 0.23$&$0.28 \pm 0.23$
             \end{tabular}\label{tab:fit_BR_t}
      \end{ruledtabular}
       \end{table}

The relative magnitude of our DE contribution is opposite to the relative magnitude of the experimentally measured branching ratios, which is also the reason why we obtain a relatively large IC contribution. This leads to a natural destructive interference phase between the IC and DE contributions. Furthermore, based on the results from the above three fitting schemes, we find that the imaginary part of the IC contribution is small. It suggests that the main contributions from the pole terms should be mainly from the intermediate ground states instead of the excited ones. This feature can be further investigated by quantitative study of the $D^0$ semileptonic decay in the NRCQM and by the charged $D$ meson decays in $D^\pm\to VP$.

\section{Summary}\label{sec:summary}
The six SCS decay channels, i.e.  $D^{0}\to \rho^{+}\pi^{-}$, $\rho^{-}\pi^{+}$, $K^{*+}K^{-}$, $K^{*-}K^{+}$, $K^{*0}\bar{K}^{0}$, and $\bar{K}^{*0}K^{0}$, have been studied within the framework of the NRCQM. We obtained the DE contributions from the NRCQM calculations and derived the IC contributions by fitting the experimental data. Since only the IC transition is involved in $D^{0}\to K^{(*)0}\bar{K}^{(*)0}$ as the leading mechanism, the combined analysis can examine the NRCQM predictions for the DE transition, and at the same time, provide crucial evidence for the role played by the IC transition. By comparing with the semileptonic decays of $D^{0}\to \rho^{-}e^{+}\nu_{e}$ and $D^{0}\to \pi^{-}e^{+}\nu_{e}$, which correspond to the weak transition form factors of $D^{0}\to \rho^-$ and $D^{0}\to \pi^-$, respectively, we show that the DE transition can be reasonably described in the NRCQM. The destructive phase between the DE and IC amplitudes is thus understandable and the contributions from the IC transition are indispensable for $D^0\to VP$. The establishment and quantitative estimate of the IC contributions should be crucial for further studies of the $D$ meson hadronic weak decays and for a better understanding of the role played by non-perturbative mechanisms in various decay channels.

\acknowledgments
This work is supported by the National Natural Science Foundation of China (Grant No. 12235018).

\bibliography{D0_to_VP.bib}
\end{document}